\documentclass[12pt]{article}

\usepackage[colorlinks=true,bookmarksopen,pdfsubject={algorithms},linkcolor={blue}, anchorcolor={black}, citecolor={blue}, filecolor={magenta}, menucolor={black}, pagecolor={red},backref=none,urlcolor={blue}]{hyperref}

\usepackage{amsmath,amssymb}
\usepackage{graphicx}
\usepackage{caption}
\usepackage{color}
\usepackage{dcolumn}
\usepackage{bm}
\usepackage{float}
\usepackage{hyperref} 
\usepackage{apacite}
\usepackage{algorithm}
\usepackage{algpseudocode}
\usepackage{subfigure}

\definecolor{background-color}{gray}{0.98}

\usepackage{mathrsfs}

\title{\color{black} Modern Monte Carlo Methods for Efficient Uncertainty Quantification and Propagation: A Survey}

\author{Jiaxin Zhang\thanks{Computer Science and Mathematics Division, Oak Ridge National Laboratory} \thanks{Email: zhangj@ornl.gov}}

\date{}
\begin{document}
\maketitle

\begin{center}
\subsubsection*{\small Article Type:}
Focus Article

\hfill \break
\thanks

\subsubsection*{Abstract}
\begin{flushleft}
{\color{black} Uncertainty quantification (UQ) includes the characterization, integration, and propagation of uncertainties that result from stochastic variations and a lack of knowledge or data in the natural world. Monte Carlo (MC) method is a sampling-based approach that has widely used for quantification and propagation of uncertainties. However, the standard MC method is often time-consuming if the simulation-based model is computationally intensive. This article gives an overview of modern MC methods to address the existing challenges of the standard MC in the context of UQ. Specifically, multilevel Monte Carlo (MLMC) extending the concept of control variates achieves a significant reduction of the computational cost by performing most evaluations with low accuracy and corresponding low cost, and relatively few evaluations at high accuracy and corresponding high cost. Multifidelity Monte Carlo (MFMC) accelerates the convergence of standard Monte Carlo by generalizing the control variates with different models having varying fidelities and varying computational costs. Multimodel Monte Carlo method (MMMC), having a different setting of MLMC and MFMC, aims to address the issue of uncertainty quantification and propagation when data for characterizing probability distributions are limited. Multimodel inference combined with importance sampling is proposed for quantifying and efficiently propagating the uncertainties resulting from small datasets. All of these three modern MC methods achieve a significant improvement of computational efficiency for probabilistic UQ, particularly uncertainty propagation. {\color{black}An algorithm summary and the corresponding code implementation are provided for each of the modern Monte Carlo methods.} The extension and application of these methods are discussed in detail. }
\end{flushleft}
\end{center}

\clearpage

\renewcommand{\baselinestretch}{1.5}
\normalsize

\clearpage

%
%

\section*{\sffamily \Large INTRODUCTION} 
Uncertainty Quantification (UQ) involves the quantitative characterization and reduction of uncertainties in the context of computational science and engineering (CSE). Both computational models and measured data, combined with theoretical analysis, are utilized in the UQ context. Practically speaking, UQ is playing an increasingly critical role in many different tasks, including model calibration, sensitivity analysis, experimental design, verification and validation, design with uncertainty, reliability analysis, risk evaluation, and decision making. Therefore, UQ benefits from many approaches and techniques in computational statistics and applied mathematics but concentrates these ideas on complex computational models and simulations. UQ has become an essential aspect of the development of CSE and widely used in many science and engineering fields \cite{ghanem2017handbook}, such as computational fluid dynamics \cite{najm2009uncertainty,le2010spectral,bijl2013uncertainty}, computational mechanics and materials \cite{soize2017uncertainty,chernatynskiy2013uncertainty,wang2020uncertainty}, structural reliability and safety \cite{marelli2014uqlab,bae2004epistemic}, environmental science \cite{oppenheimer2016expert}, chemical science \cite{najm2009uncertainty,ryu2019bayesian}, etc. 

{\color{black}
Although many sources of uncertainty exist, they are typically categorized as \emph{aleatory}, resulting from intrinsic randomness or variability, or \emph{epistemic}, resulting from a lack of complete knowledge (or data). Most problems of engineering and science interest involve both types of uncertainties \cite{smith2013uncertainty,sullivan2015introduction,soize2017uncertainty,ghanem2017handbook}. In many cases, it is challenging to determine whether a particular uncertainty should be put in the aleatory category or the epistemic category \cite{der2009aleatory}.} In recent years, UQ has gained popularity as an essential approach to assess the effect of variability, randomness and lack of knowledge on the response output, i.e., the quantity of interest (QoI) \cite{ghanem2017handbook}. This program is referred to as a forward UQ problem (also called uncertainty propagation). Probabilistic methods can be readily applied by converting the source of uncertainties (input data uncertainty, initial and boundary uncertainty, physical model uncertainty, model parameter uncertainty, etc.) into random variables of fields. Methods in probabilistic UQ framework can be typically categorized into two groups: non-sampling methods and sampling-based methods. A typical non-sampling method is the stochastic Galerkin method \cite{ghanem2003stochastic}, which is based on a representation of the uncertainty solution as a polynomial expansion. This method is accurate and allows for a large number of uncertainties, but it is highly intrusive. An example of sampling-based methods is the stochastic collocation method \cite{babuvska2007stochastic}, which samples a stochastic PDE at specific collocation points in the stochastic space with an interpolating polynomial, resulting in a non-intrusive scheme. However, the stochastic collocation method suffers from a similar challenge with the stochastic Galerkin approach, which is the ``{\em curse of dimensionality''}. 

Monte Carlo (MC) method is one of the sampling-based approaches which can handle the issue of high dimensionality \cite{kalos2009monte} in the probabilistic UQ framework. MC method has many advantages, such as non-intrusive, robust, flexible, and simple for implementation. However, an obvious drawback is its slow convergence rate of $ \mathcal{O}(N^{-1/2})$, where $N$ is the number of function evaluations. In other words, it is very time-consuming to converge if the simulation-based model is computationally intensive. Many efforts have been made to reduce the computational cost required to obtain accurate statistics. Conventionally, a variety of ways are proposed to accelerate the convergence rate of the standard MC method. Typically, methods based on variance reduction techniques \cite{rubinstein2016simulation} are relatively widespread used, such as stratified sampling, Latin hypercube sampling, control variates, importance sampling, etc. An alternative method is quasi-Monte Carlo (QMC) \cite{caflisch1998monte}, which uses low-discrepancy sequences, such as Sobol sequence or Halton sequence, whereas the standard MC method uses a pseudorandom sequence. QMC has a rate of convergence close to $ \mathcal{O}(N^{-1})$, which is faster than the standard MC method. These improved MC methods increase the precision of estimates that can be obtained for a given simulation or computational effort. Recently, several modern MC methods are proposed and widely applied to overcome the computational challenges in the context of uncertainty quantification and propagation. For instance, multilevel Monte Carlo methods (MLMC), developed by \cite{heinrich1998monte, giles2008multilevel}, significantly reduce the computational cost by performing most simulations with low accuracy at a corresponding low cost, with few model evaluations being achieved at a high cost. Based on the multifidelity method proposed by \cite{ng2014multifidelity}, Multifidelity Monte Carlo methods (MFMC) \cite{peherstorfer2016optimal} that combine outputs from computationally cheap low-fidelity models with output from high-fidelity models, can lead to a significant reduction of the time cost, and provide unbiased estimators of the statistics of the high-fidelity model outputs. Zhang and Shields \cite{zhang2018quantification} proposed a multimodel Monte Carlo method (MMMC) for quantifying and efficiently propagating the uncertainties resulting from small datasets. This method utilizing multimodel inference and importance sampling achieves a significant cost reduction by collapsing the multiple MC loops to a single MC loop in the forward uncertainty propagation.

In this study, we focus on the review of modern Monte Carlo methods in the context of UQ, particularly uncertainty propagation. These modern MC methods are all extensions of the standard MC method but either reduce the computational cost or show the faster convergence rate of the response estimators. MLMC utilizes the control variates to reallocate the overall computational budget among different hierarchical levels of simulations according to the number of samples required to decrease the variance at each level. MFMC, as similar to MLMC, is inspired by control variates but uses more general low-fidelity models with properties that cannot necessarily be well described by rates. MMMC, differing from the above two methods, integrates multimodel inference and importance sampling to address the computational challenge of the propagation of imprecise probabilities caused by a lack of data. Also, we illustrate how to apply these modern Monte Carlo methods to achieve efficient and accurate statistical estimates in the probabilistic UQ framework. 

The paper is structured as follows. In section 2, we provide a brief overview of the UQ concepts and the standard MC method. The application of the MC method to uncertainty propagation is mainly introduced. Section 3 shows a review of the modern MC methods, i.e., MLMC, MFMC, and MMMC. We formulate the mathematical setting, algorithm procedure, and describe their corresponding applications in the context of UQ. Finally, we conclude with a discussion in Section 4 about the next frontiers for future research in this area.

\section*{\sffamily \Large \MakeUppercase{Standard Monte Carlo Method for UQ} }

Most of the predictions that are necessary for decision making in science and engineering are made based on computational models. These models are assumed to approximate reality, but they always have uncertainties in their predictions due to many sources of variability, randomness, and stochasticity. The response of a model is also subjective to variability due to uncertainties on model parameters, which may be due to measurement errors, possible wrong assumptions, and other factors. Probability theory is one of the essential ways to consider these uncertainties by describing the uncertain parameters as random variables, random processes, or/and random fields. This approach allows us to quantify the variability of the response in a probabilistic way. 

\begin{figure}[H]
	\centering
	\includegraphics[width=5in]{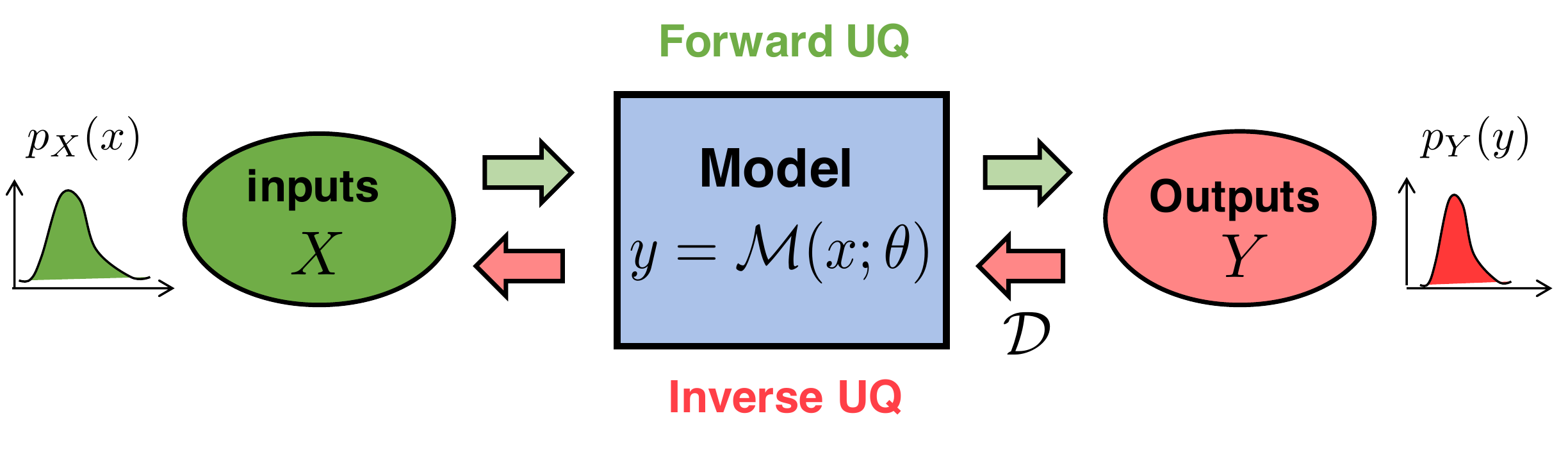}
	\caption{Illustration of Uncertainty Quantification (UQ): forward UQ and inverse UQ }
	\label{fig0} 
\end{figure}

In the probabilistic UQ framework, there are two major types of UQ problems: {\em forward UQ} and {\em inverse UQ}, as shown in Figure \ref{fig0}. Consider a computational model $\mathcal{M}$: $\mathcal{X} \rightarrow \mathcal{Y}$, and let the uncertainties in the inputs be represented by a random variable, ${X} \in \mathcal{X}$, with probability density function $p_X(x)$. The random variable representing the output is $Y=\mathcal{M}(X)$ and $\mathcal{M}$ is a real-valued function defined over $\mathcal{X}$. The goal of forward UQ (uncertainty propagation) is to estimate statistics of the outputs represented by a random variable $Y \in \mathcal{Y}$, e.g., the probability distribution $p_Y(y)$, the expectation, 
\begin{equation}
\mathbb{E}[\mathcal{M}] = \int_{\mathcal{X}}\mathcal{M}(\bm{x}) p (\bm{x})d\bm{x},
\end{equation}
and the variance,
\begin{equation}
\mathbb{V}[\mathcal{M}] = \mathbb{E}[\mathcal{M}^2] - \mathbb{E}[\mathcal{M}]^2,
\end{equation}
where we assume to exist. On the other hand, inverse UQ means that an inverse assessment of model parameter uncertainty $p(\theta | \mathcal{D})$ given measured outputs for specific model inputs $\mathcal{D} =\left\{x_i, y_i \right\}$. In this paper, we mainly focus on the forward UQ, that is the propagation of uncertainties of the input random variable ${X}$ through the computational model $\mathcal{M}$. 

MC method, commonly used for uncertainty propagation, generates several samples of the input random variable according to its probability distribution. Each of these samples defines a deterministic problem, which is solved by a deterministic technique, for example, a simulation or computational model, and finally generates an amount of output data. All of these output data are then combined to assess the variability of the random system statistically. If a large number of samples are performed, the MC method can achieve a complete description of the statistical behavior of the random system. 

{\color{black}
In the mathematical setting, MC method approximates the expected value of a random variable $Y$, such as $s = \mathbb{E} [Y]$.  In general, we generate values {\color{black}$y_1,...,y_n$} independently and randomly from the distribution of $Y$ and take their average 
\begin{equation}
{\color{black}
\hat{s}_n = \frac{1}{n}\sum_{i=1}^n y_i \label{eq: mc_1}}
\end{equation}
as the estimate of $s$. 
In the setting of independent and identically distributed (i.i.d.) sampling, $\hat{s}_n$ is a random variable and the mean of $\hat{s}_n$ is 
\begin{equation}
{\color{black}
\mathbb{E}[\hat{s}_n] = \frac{1}{n}\sum_{i=1}^n \mathbb{E}[y_i] = s.}
\end{equation}
The MC estimator is an unbiased estimator since the expected value of $\hat{s}_n$ is equal to $s$.  According to the {\em strong law of large numbers}, the average converges almost surely to the true expected value:
\begin{equation}
\mathbb{P}(\lim_{n \rightarrow \infty} |\hat{s}_n-s|=0)=1,
\end{equation}
provided that the variance of the individual terms, $\mathbb{V}[Y]$ is bounded \cite{rubinstein2016simulation}. 
Suppose that $\mathbb{V}[Y] = \sigma^2 < \infty$, we have 
\begin{equation}
\mathbb{V}[\hat{s}_n] =  \mathbb{E}[(s-\hat{s}_n)^2] = \frac{\sigma^2}{n},
\end{equation}
and the root mean square error (RMSE) of $\hat{s}_n$ is $ {\sigma}/{\sqrt{n}}$, which means the error of MC estimator is of order $n^{-1/2}$ and we typically write $\textup{RMSE}=\mathcal{O}(n^{-1/2})$, where $\mathcal{O}(\cdot)$ notation allows us to focus on the rates of convergence as $n \rightarrow \infty$.

One of the strengths of the MC method is that the sample values can be used for error estimation \cite{mcbook}. The most commonly used estimates of $\sigma^2$ are 
\begin{equation}
{\color{black}\zeta^2 = \frac{1}{n-1}\sum_{i=1}^n (y_i-\hat{s}_n)^2,   \label{eq:mc_err}}
\end{equation} 
\begin{equation}
{\color{black} \hat{\zeta}^2 = \frac{1}{n}\sum_{i=1}^n (y_i-\hat{s}_n)^2,}
 \end{equation} 
and the estimator in Eq.\eqref{eq:mc_err} is unbiased since $\mathbb{E}[\zeta^2 ]=\sigma^2$ for $n \ge 2$.  From the {\em central limit theorem}, the error $\hat{s}_n-s$ has approximately a normal distribution with mean 0 and variance $\sigma^2/n$. Therefore, we can estimate confidence intervals around the estimate $\hat{s}_n$. Summary \ref{algo:MC1} provides a brief summary of using the standard MC method for uncertainty propagation.

}

{\color{black}
\makeatletter
\newenvironment{megaalgorithm}[1][htb]{%
  \renewcommand{\ALG@name}{Summary}
  \begin{algorithm}[#1]%
 }{\end{algorithm}}
\makeatother

\begin{megaalgorithm}
\begin{algorithmic}[1]
\State Identify the input random variable ${X}$ from the computational model $\mathcal{M}$
\State Determine the variability of ${X}$ by assuming a probability density function $p_X(x)$
\State Generate $n \in \mathbb{N}$ i.i.d. samples from $p_X(x)$, $\left\{ {x}_{1},...,{x}_{N}\right\} \sim p_X(x)$
\For{$i = 1,2,...,n$}
\State Compute response output values through the model evaluations ${y}_{i} = \mathcal{M}({x}_{i})$
\EndFor
\State Compute the statistics of the output random variable $Y$
\end{algorithmic}
\caption{Standard Monte Carlo method for uncertainty propagation}
\label{algo:MC1}
\end{megaalgorithm}
}


The MC method does not require a new computer code to simulate a random or stochastic simulation model. If the computational model is available, the stochastic simulation can be performed by running the model several times, changing only the input random variable that is randomly drawn by a specific probability distribution. This non-intrusive characteristic is an excellent advantage of the MC method when compared with other methods for uncertainty propagation. However, standard MC is a very time-consuming method, which makes its use unfeasible for complex high-fidelity simulations. The reason is that the standard MC often needs extensive model evaluations to obtain an accurate approximation. Moreover, the MC method relies on the ability to sample from the assumed probability distribution easily, but doing so is not always possible. 

Many efforts have been made to reduce the computational cost required to achieve accurate MC statistics. A typical way is to reduce the variance of the estimates. This method is also called variance reduction (VR) techniques, which typically include common random numbers, stratified sampling, importance sampling, control variates, antithetic variates, etc. In addition, there are also a series of advanced VR methods, for example, Latin hypercube sampling \cite{stein1987large, shields2016generalization}, adaptive importance sampling \cite{cappe2008adaptive, cornuet2012adaptive}, sequential Monte Carlo methods \cite{doucet2000sequential, liu1998sequential}, which are generalizations, extensions or hybrids of the basic VR methods. Another typical idea is the quasi-Monte Carlo method \cite{niederreiter1978quasi, niederreiter1992random,caflisch1998monte}, which is to choose deterministic sample points that are spread out as uniformly as mathematically possible. Moreover, when it is not feasible to sample from the probability distribution $p$, an alternative is to use importance sampling, but a more general approach is Markov chain Monte Carlo (MCMC) methods, which form a sequence of estimators that converge toward the target probability density. Metropolis algorithm, proposed by Metropolis \cite{metropolis1953equation} and generalized by Hastings \cite{hastings1970monte}, and Gibbs sampling, introduced by German and German \cite{geman1984stochastic}, are two most foundational algorithms in the MCMC research area. MCMC methods combined with Bayesian inference are widely applied for the estimation of model parameters in inverse UQ problems.  

Recently, several new ideas from the UQ community are developed to address the computational challenges in applying MC-based methods for uncertainty propagation. By extending the concept of control variates, multilevel Monte Carlo (MLMC) and multifidelity Monte Carlo (MFMC) are proposed by leveraging the model hierarchy with an optimal balance between the variance and bias estimates. Both methods aim to provide more accurate MC estimates given a specific computational budget. Additionally, as discussed before, the probability distribution $p$ of random input has a critical impact on the response output. However, in practice, $p$ is often assumed subjectively given specific knowledge or information. It is difficult to identify a precise and objective $p$ if only sparse or limited data is observed. This, therefore, leads to an issue of {\em imprecise probability} \cite{walley2000towards,augustin2014introduction}. Multimodel Monte Carlo (MMMC) method addresses this issue by combining multimodel inference and importance sampling and thus achieves effective quantification and efficient propagation of imprecise probabilities resulting from small datasets. This paper aims to give an overview of these three modern Monte Carlo methods for uncertainty quantification and propagation.

\section*{\sffamily \Large \MakeUppercase{Modern Monte Carlo Methods for UQ} }
{\color{black}This section focuses on three modern Monte Carlo methods: multilevel Monte Carlo methods, multifidelity Monte Carlo methods, and multimodel Monte Carlo methods. For each MC method, we provide a brief review of the algorithm procedure, generalization, and extension of original methods, and finally followed by the applications to uncertainty quantification and propagation.  }                                                                                                                                                                                                                                                                                                                                                                                                                                                                                            

\subsection*{\sffamily \large {Multilevel Monte Carlo Methods} }

Multilevel Monte Carlo (MLMC) is a recently proposed method that makes use of a control variates technique to dramatically reduces the computational cost by performing relatively few high-fidelity simulations at a high cost, with most low-fidelity simulations at a corresponding low cost \cite{giles2015multilevel}. The core idea of MLMC is to reallocate the overall computational cost among different hierarchical levels of simulations according to the number of samples required to decrease the variance at each level \cite{giles2013multilevel, giles2015multilevel}. Heinrich was the first one to apply the MLMC for the parametric integration and evaluation of functionals from the solution of integral equations \cite{heinrich1998monte, heinrich2000multilevel, heinrich2001multilevel}. Using a similar idea, Kebaier \cite{kebaier2005statistical} proposed a two-level MC method to approximately solve the stochastic differential equations (SDEs). Giles \cite{giles2008multilevel} then generalized and applied MLMC in the context of SDEs for option pricing. A complete review of MLMC methods can be found by \cite{giles2015multilevel}. 

\subsubsection*{\sffamily \normalsize Control Variates Techniques}  
\label{sec:cv}      
{\color{black}                                                                                                                   
Control variates is one of the classic variance reduction techniques in MC method \cite{glasserman2013monte}. Assuming that we are interested in an expectation estimator $\mathbb{E}[\mathcal{M}]$ given a control variate $\mathcal{G}$ that is correlated to $\mathcal{M}$ and has a known expectation $\mathbb{E}[\mathcal{G}]$, then an unbiased estimator $\hat{s}_N^{cv}$ for $\mathbb{E}[\mathcal{M}]$ based on $N$ i.i.d. realizations $\bm{x}_1,...,\bm{x}_N \in \mathcal{X}$ of the random variable $X$ is
\begin{equation}
\hat{s}_N^{cv} = \frac{1}{N}\sum_{i=1}^N \left[ \mathcal{M}(\bm{x}_i) - \lambda (\mathcal{G}(\bm{x}_i) - {\color{black}\mathbb{E}[\mathcal{G}]}) \right].
\end{equation}
The optimal value of $\lambda$ is {\color{black} $\rho \sqrt{\mathbb{V}[\mathcal{M}] / \mathbb{V}[\mathcal{G}]}$ }, where $\rho$ is the correlation between $\mathcal{M}$ and $\mathcal{G}$. Compared to the variance of the standard MC estimator $\mathbb{V}[\hat{s}_N]$, the variance of the control variates estimator is 
\begin{equation}
\mathbb{V}[\hat{s}_{N}^{cv}] = (1-\rho^2) \mathbb{V}[\hat{s}_N]
\label{eq:mf_00}
\end{equation}
where $| \rho|<1$ is the correlation coefficient. Eq. \eqref{eq:mf_00} shows that the variance reduction is strongly determined by the correlation degree between $\mathcal{M}$ and $\mathcal{G}$. 

It is not difficult to extend the control variates to a two-level MC estimator. Considering a computationally cheaper model $\mathcal{M}^{(0)}$ which is correlated to a model $\mathcal{M}^{(1)}$, then we can use a unbiased two-level estimator to estimate $\mathbb{E}[\mathcal{M}^{(1)}]$ 
\begin{equation}
{\color{black}
\mathbb{E}[\mathcal{M}^{(1)}] = \frac{1}{N_0}\sum_{i=1}^{N_0}\mathcal{M}^{(0)}(\bm{x}_i^{(0)}) + \frac{1}{N_1}\sum_{i=1}^{N_1}\left(\mathcal{M}^{(1)}(\bm{x}_i^{(1)}) - \mathcal{M}^{(0)}(\bm{x}_i^{(1)}) \right)
}
\end{equation}
since 
\begin{equation}
\mathbb{E}[\mathcal{M}^{(1)}] = \mathbb{E}[\mathcal{M}^{(0)}] + \mathbb{E}[\mathcal{M}^{(1)}-\mathcal{M}^{(0)}].
\end{equation}
Note that there are two critical differences from the standard control variates: we use $\lambda = 1$ and $\mathbb{E}[\mathcal{M}^{(0)}]$ is unknown so it needs to be estimated. 

Let $V^{(0)}$ and $V^{(1)}$ to be the variance of $\mathcal{M}^{(0)}$ and $\mathcal{M}^{(1)} - \mathcal{M}^{(0)}$, $C^{(0)}$ and $C^{(1)}$ to be the cost of single realization of $\mathcal{M}^{(0)}$ and $\mathcal{M}^{(1)} - \mathcal{M}^{(0)}$ respectively, then the overall variance $V$ and total cost $C$ are 
\begin{equation}
V = \frac{V^{(0)}}{N_0} + \frac{V^{(1)}}{N_1}, \quad C= {C^{(0)}}{N_0} + {C^{(1)}}{N_1}.
\end{equation}

The parameters $N_0$ and $N_1$ are chosen such that the overall variance is minimized for a given computational budget $C^* \in \mathbb{R}_{+}$. The solution to the optimization problem using Lagrange multiplier gives
\begin{equation}
\frac{N_1}{N_0} = \frac{\sqrt{V^{(1)}/C^{(1)}}}{\sqrt{V^{(0)}/C^{(0)}}}.
\end{equation}

\subsubsection*{\sffamily \normalsize Multilevel Monte Carlo Algorithm} 

It is quite straightforward to generalize this two-level MC method to multilevel MC algorithm. Consider a sequence simulation $\left\{ \mathcal{M}^{(\ell)}: \ell = 0,...,L \right\}$ with increasing accuracy but also an increasing computational cost, 
\begin{equation}
{\color{black} \mathcal{M}^{(0)}, \mathcal{M}^{(1)}, \cdots ,\mathcal{M}^{(L)} = \mathcal{M}.}
\end{equation}
According to the expectation identity
\begin{equation}
\mathbb{E}[\mathcal{M}^{(L)}] = \mathbb{E}[\mathcal{M}^{(0)}] + \sum_{\ell =1 }^{L}\mathbb{E}[\mathcal{M}^{(\ell)} - \mathcal{M}^{(\ell-1)}],
\end{equation}
we have the unbiased estimator for $\mathbb{E}[\mathcal{M}^{(L)}]$ by means of correction with respect to the next lower level,
\begin{equation}
{\color{black}
\hat{s}^{ml} = \frac{1}{N_0}\sum_{i=1}^{N_0}\mathcal{M}^{(0)}(\bm{x}_i^{(0)}) + \sum_{\ell=1}^L \left\{ \frac{1}{N_{\ell}} \sum_{i = 1}^{N_{\ell}} \left[ \mathcal{M}^{(\ell)}(\bm{x}_i^{(\ell)}) - \mathcal{M}^{(\ell-1)}(\bm{x}_i^{(\ell)}) \right] \right\},
\label{eq:ml2}
}
\end{equation}
where $\bm{x}_i^{(\ell)}$ are independent samples that are used at each level of correction. 

Then we focus on the numerical cost and variance of the MLMC estimator. Similarly, we define $V^{(0)}$ and $C^{(0)}$ to be the variance and cost of one sample of $\mathcal{M}^{(0)}$, and $V^{(\ell)}$ and $C^{(\ell)}$ to be the variance and cost of one sample of $\mathcal{M}^{(\ell)}- \mathcal{M}^{(\ell-1)}$, then the overall variance and cost of the multilevel MC estimator is 
\begin{equation}
V = \sum_{\ell=0}^L N_{\ell}^{-1}V^{(\ell)}, \quad C= \sum_{\ell=0}^L N_{\ell}C^{(\ell)}. 
\end{equation}

{\color{black} It is important to determine the ideal number of samples for per level}. Using a Lagrange multiplier $\xi^2$ to minimize the cost for a fixed variance 
\begin{equation}
\mathscr{L} = \sum_{\ell=0}^{L} (N_{\ell}C^{(\ell)} + \xi^2 N_{\ell}^{-1}V^{(\ell)} ) \Longrightarrow \frac{\partial \mathscr{L}}{\partial N_{\ell}} =0
\end{equation}
yields $N_{\ell} = \xi \sqrt{V^{(\ell)}/C^{(\ell)}}$. Setting the total variance equal to $\varepsilon^2$ gives $\xi =\varepsilon^{-2} \sum_{\ell=0}^{L}\sqrt{V^{(\ell)} C^{(\ell)}}$, the total computational cost is thus 
\begin{equation}
C_{ml} = \sum_{\ell=0}^{L} N_{\ell}C^{(\ell)} = \varepsilon^{-2} \left( \sum_{\ell=0}^L \sqrt{V^{(\ell)}C^{(\ell)}}\right)
\end{equation}
in contrast to the standard MC cost which is approximately $\varepsilon^{-2}V^{(0)}C^{(L)}$. 

It is necessary to note that the MLMC cost is reduced by a factor of $V^{(L)}/V^{(0)}$ if $\sqrt{V^{(\ell)}C^{(\ell)}}$ increases with level $\ell$ or it is reduced by a factor of $C^{(0)}/C^{(L)}$ if $\sqrt{V^{(\ell)}C^{(\ell)}}$ decreases with level $\ell$. The MLMC algorithm for uncertainty propagation is briefly summarized as {\color{black} Summary \ref{algo:MC2}}. Interested readers can find more discussions, particularly about the convergence test of MLMC from \cite{giles2015multilevel}. } {\color{black} The code implementation of MLMC for forward UQ can be found at \url{https://github.com/NASA/MLMCPy}. {\textit{MLMCPy}} is an open source Python implementation of the Multilevel Monte Carlo (MLMC) method for uncertainty propagation.}

{\color{black}
\makeatletter
\newenvironment{megaalgorithm}[1][htb]{%
  \renewcommand{\ALG@name}{Summary}
  \begin{algorithm}[#1]%
 }{\end{algorithm}}
\makeatother

\begin{megaalgorithm}
\begin{algorithmic}[1]
\State Define initial target of $N_0$ samples on levels $\ell = 0,1,2$
\While {additional samples are required}
\State Evaluate model $\mathcal{M^{(\ell)}}$ at additional samples on all levels, $\ell = 0,...,L$
\State Compute the sample variance $V^{(\ell)}$ on all levels, $\ell = 0,...,L$
\State Determine the optimal number of samples $N_{\ell}$ on all levels, $\ell = 0,...,L$, as 
$$ N_{\ell} = \varepsilon^{-2} \sqrt{V^{(\ell)}/C^{(\ell)}} \left( \sum_{\ell=0}^{L}\sqrt{V^{(\ell)} C^{(\ell)}} \right) $$
\If {$L \ge 2$} test for convergence \EndIf
\If {not converged} $L = L+1$
\EndIf
\EndWhile
\State Calculate the multilevel estimator $\hat{s}^{ml}$ as in Eq. \eqref{eq:ml2}
\end{algorithmic}
\caption{Multilevel Monte Carlo methods for uncertainty propagation}
\label{algo:MC2}
\end{megaalgorithm}
}

{\color{black}
\subsubsection*{\sffamily \normalsize Multilevel Monte Carlo extensions and applications}  
An interesting extension of MLMC is the randomized multilevel Monte Carlo method, which is proposed by Rhee and Glynn \cite{rhee2015unbiased}. This method uses $n$ total sample but for each sample, it performs a simulation on level $\ell$ with probability $p_{\ell}$, rather than using the optimal number of samples on each level based on the estimate of the variance. Another significant extension of the MLMC method is the multi-index Monte Carlo (MIMC) method, developed by \cite{haji2016multi}. MIMC method generalizes the level from one-dimension to multiple directions and thus performs a vector of integer indices $\ell$ instead of the scalar level index $\ell$ in MLMC. Many recent studies are developed based on the key aspect of MIMC. Haji-Ali and Tempone \cite{haji2018multilevel} proposed a hybrid multilevel and multi-index Monte Carlo methods and apply them for the McKean-Vlasov equation. The multi-index method can also be extended to the Markov chain Monte Carlo (MCMC) method \cite{jasra2018multi}. It is also necessary to introduce a variant of MLMC, which uses Quasi-Monte Carlo (QMC) samples replacing i.i.d. Monte Carlo random samples \cite{giles2009multilevel}. There have been some studies on the theoretical foundations and practical applications for multilevel QMC methods \cite{scheichl2017quasi, kuo2017multilevel, herrmann2019multilevel}. Many efforts have been also made to combine multilevel methods with MCMC. For example, Dodwell et al.\cite{dodwell2015hierarchical, dodwell2019multilevel} developed a hierarchical multilevel Markov chain Monte Carlo method to address the problem of the prohibitively large computational cost of existing MCMC methods for large-scale applications with high-dimensional parameter spaces, e.g., in uncertainty quantification in porous media flow. Several researchers also investigate to combine the multilevel method with data assimilation \cite{beskos2017multilevel}, for example, particle filters \cite{jasra2017multilevel} and Kalman filtering \cite{hoel2016multilevel}.  

Typically, the MLMC method is used for applications in finance, e.g. SDEs and SPDEs \cite{giles2012stochastic} but recently the application of MLMC to UQ problems has attracted an increasing attention \cite{teckentrup2013multilevel}. Jasra et al. \cite{jasra2016forward} conducted forward and inverse UQ using MLMC and multilevel sequential Monte Carlo (MLSMC) sampling algorithms for an elliptic nonlocal equation.  Eigel et al. \cite{eigel2016adaptive} introduced an adaptive MLMC with stochastic bounds for quantities of interest with uncertain data. Elfverson et al. \cite{elfverson2016multilevel} proposed to use MLMC methods for computing failure probabilities of systems modeled as numerical deterministic models with uncertain input data. Pisaroni et al. \cite{pisaroni2017continuation} proposed a continuation MLMC method for uncertainty quantification in compressible inviscid aerodynamics and quantify the uncertain system outputs using the MLMC method with central moment estimation \cite{pisaroni2017quantifying}. Fairbanks et al. \cite{fairbanks2017low} presented an extension of MLMC, referred to as multilevel control variates (MLCV), where a low-rank approximation to the solution on each grid is used as a control variate for estimating the expectations of high-dimensional uncertain systems.  Ali et al. \cite{ali2017multilevel} provided an MLMC analysis for optimal control of elliptic PDEs with random coefficients, which is motivated by the need to study the impact of data uncertainties and material imperfections on the solution to optimal control problems.  Scheichl et al. \cite{scheichl2017quasi} focused on estimating Bayesian posterior expectations in elliptic inverse problems using quasi-MC and MLMC method.  Rey et al. \cite{rey2019quantifying} used the MLMC method to effectively quantify the uncertainties in contact mechanics of rough surfaces. 

}

\subsection*{\sffamily \large {Multifidelity Monte Carlo Methods} }
{\color{black} Multifidelity methods make use of multiple approximate models and other sources of knowledge to accelerate the time-consuming tasks, for example, UQ, design optimization and statistical inference.} Multifidelity Monte Carlo (MFMC) aims to accelerate the statistical estimation by combining the outputs from the high-fidelity model and a large number of low-fidelity models. Rather than just replacing the high-fidelity model with a low-fidelity model, MFMC utilizes recourses to the high-fidelity model to establish convergence and accuracy guarantees on the statistical quantities of the output response. In 2014, Ng and Willcox \cite{ng2014multifidelity} developed a multifidelity method to estimate the mean using the control variates technique. This work utilized efficient low-fidelity models to reduce the computational cost of high-fidelity simulation in optimization with uncertainty. Followed by this work, Peherstorfer \cite{peherstorfer2016optimal, peherstorfer2018survey} extended the multifidelity method by utilizing a large number of low-fidelity models and performed optimal management of model allocations based on their relative fidelities and costs. In fact, several different types of low-fidelity models, for example, projection-based models \cite{benner2015survey, peherstorfer2016data, swischuk2019projection}, surrogate models \cite{park2017remarks}, up-scaled models \cite{durlofsky2012uncertainty} could be applied to the MFMC framework. 

\subsubsection*{\sffamily \normalsize Control variates in multifidelity estimator}  
The use of control variates technique is to reduce the estimator variance by utilizing the correlation with an auxiliary random variable. As discussed before, the statistics of the auxiliary random variable are known in the standard control variates approach. This requirement can be relaxed by estimating the statistics of the auxiliary random variable from prior knowledge \cite{pasupathy2012control}. Multifidelity method aims to construct effective auxiliary random variables from low-fidelity models. 

{\color{black} Let $\mathcal{M}_{\text{hi}}$ be the high-fidelity model and} $\mathcal{M}_{\text{lo}}^{(1)}, ..., \mathcal{M}_{\text{lo}}^{(k)}, k \in \mathbb{N}$ be the low-fidelity models. Multifidelity methods use the random variables $\mathcal{M}_{\text{lo}}^{(1)}(X), ..., \mathcal{M}_{\text{lo}}^{(k)}(X) $ stemming from the low-fidelity models as control variates to estimate the statistics of the random variable $\mathcal{M}_{\text{hi}}(X)$ of the high-fidelity model. Considering $n_0 \in \mathbb{N}$ to be the number of high-fidelity model evaluations, and $n_i \in \mathbb{N}$ to be the number of low-fidelity models $\mathcal{M}_{\text{lo}}^{(i)}$ for $i=1,...,k$, where $0<n_0 \le n_1 \le \cdots \le n_k$, we draw $n_k$ realizations $\bm{x}_1,...,\bm{x}_{n_k}$ from the random variable $X$ and evaluate the high-fidelity model outputs $\mathcal{M}_{\text{hi}}(\bm{x}_1), ..., \mathcal{M}_{\text{hi}}(\bm{x}_{n_0})$ and low-fidelity outputs $\mathcal{M}_{\text{lo}}^{(i)}(\bm{x}_1), ..., \mathcal{M}_{\text{lo}}^{(i)}(\bm{x}_{n_i})$ for $i=1,..., k$. Therefore, the MC estimates can be derived by using all of these model outputs: 
\begin{equation}
\hat{s} _{n_0}^{\text{hi}} = \frac{1}{n_0}\sum_{j=1}^{n_0}\mathcal{M}_{\text{hi}}(\bm{x}_j), \quad \hat{s}_{n_i}^{(i)} = \frac{1}{n_i}\sum_{j=1}^{n_i}\mathcal{M}_{\text{lo}}^{(i)}(\bm{x}_j), \quad i = 1,...,k.
\label{eq:mf_0}
\end{equation}
The multifidelity estimator of $\mathbb{E}[\mathcal{M}_{\text{hi}}]$ is thus \cite{peherstorfer2016optimal}
\begin{equation}
\hat{s}^{mf} = \hat{s} _{n_0}^{\text{hi}} + \sum_{i=1}^k \beta_i \left( \hat{s}_{n_i}^{(i)} - \hat{s}_{n_{i-1}}^{(i)} \right) \label{eq:mf_1}.
\end{equation}
It is noted that the computation of $\hat{s}_{n_{i-1}}^{(i)}$ in Eq.~\eqref{eq:mf_1} re-uses the first $n_{i-1}$ model outputs $\mathcal{M}_{\text{lo}}^{(i)}(\bm{x}_1),...,\mathcal{M}_{\text{lo}}^{(i)}(\bm{x}_{n_{i-1}})$ of the $n_i$ model outputs generated to compute $\hat{s}_{n_i}^{(i)}$ in Eq.~\eqref{eq:mf_0}. 
The control variate coefficients $\beta_1,...,\beta_k \in \mathbb{R}$ balance the terms $\hat{s}_{n_i}^{(i)} - \hat{s}_{n_{i-1}}^{(i)} $ stemming from the low-fidelity models and the term $\hat{s} _{n_0}^{\text{hi}}$ from the high-fidelity model. 

The multifidelity estimator $\hat{s}^{mf}$ is an unbiased estimator of $\mathbb{E}[\mathcal{M}_{\text{hi}}]$ since 
\begin{equation}
\mathbb{E}[\hat{s}^{mf}] = \mathbb{E}[\hat{s} _{n_0}^{\text{hi}}] + \sum_{i=1}^k \beta_i \mathbb{E} [\hat{s}_{n_i}^{(i)} - \hat{s}_{n_{i-1}}^{(i)}] =\mathbb{E}[\mathcal{M}_{\text{hi}}].
\end{equation}
Thus, the mean square error (MSE) of the estimator $\hat{s}^{mf}$ is equal to the variance $\mathbb{V}[\hat{s}^{mf}] $ of the estimator, $e(\hat{s}^{mf}) = \mathbb{V}[\hat{s}^{mf}] $.
%
%
The variance $\mathbb{V}[\hat{s}^{mf}]$ of the multifidelity estimator $\hat{s}^{mf}$ is 
\begin{equation}
\mathbb{V}[\hat{s}^{mf}] = \frac{\sigma_{\text{hi}}^2}{n_0} + \sum_{i=1}^k \left( \frac{1}{n_{i-1}} - \frac{1}{n_i} \right) (\beta_i^2 \sigma_i^2 - 2 \beta_i \rho_i \sigma_{\text{hi}} \sigma_i),
\end{equation}
where 
\begin{equation}
\sigma_{\text{hi}}^2 = \mathbb{V}[\mathcal{M}_{\text{hi}}], \quad \sigma_i^2 = \mathbb{V}[\mathcal{M}_{\text{lo}}^{(i)}], \quad i = 1,...,k.
\end{equation}
are the variance of $\mathcal{M}_{\text{hi}} (X)$ and $\mathcal{M}_{\text{lo}}^{(i)} (X)$ respectively. $|\rho_i| \le 1$ is the Pearson correlation coefficient of the random variables $\mathcal{M}_{\text{hi}}(X)$ and $\mathcal{M}_{\text{lo}}^{(i)}(X)$ for $i=1,...,k$. 

\subsubsection*{\sffamily \normalsize Multifidelity Monte Carlo algorithm}  

The computational cost $C_{mf}$ of the multifidelity estimator $\hat{s}^{mf}$ in Eq. \eqref{eq:mf_1} has a critical impact on the performance of MFMC algorithm. $C_{mf}$ depends on the number of model evaluations and the single cost of each evaluation. It is therefore given by
\begin{equation}
C_{mf} = n_0c_{\text{hi}} + \sum_{i=1}^k n_i c_{\text{lo}}^{(i)} = \bm{n}^{T}\bm{c}
\end{equation}
where $\bm{n} = [n_0, ...,n_k]^T$ and $\bm{c} = [c_{\text{hi}}, c_{\text{lo}}^{(1)},...,c_{\text{lo}}^{(k)}]^T$. The high-fidelity model $\mathcal{M}_{\text{hi}}$ is evaluated at $n_0$ realizations and the low-fidelity model $\mathcal{M}_{\text{lo}}^{(i)}$ is evaluated at $n_i$ realizations of $X$ for $i=1,.,,,k$. 

To achieve efficient propagation of uncertainty, MFMC aims to minimize the variance $ \mathbb{V}[\hat{s}^{mf}] $ of the multifidelity estimator $\hat{s}^{mf}$ given a computational budget $C^* \in \mathbb{R}_{+}$. This goal can be achieved by solving a optimization problem
\begin{equation}
\begin{aligned}
&\underset{n_0, n_1,...,n_k, \ \beta_1, \beta_2,...,\beta_k} {\text{minimize}} 
&&\mathbb{V}[\hat{s}^{mf}] \\
& \quad \quad \text{subject to}
&& n_0 >0, \\ 
&&& n_i \ge n_{i-1}, \quad i= 1,...,k, \\
&&& C_{mf} \le C^*
\end{aligned}
\label{eq:opt_mf}
\end{equation}
where the control variate coefficient $\beta_1, ...\beta_k$ and the number of model evaluations $n_0,....,n_k$ are design variables. The constraints are $0 < n_0 \le n_1 \le,...,\le n_k$ and the cost $C_{mf}$ is less than the computational budget $C^*$.
 

Peherstorfer et al.\cite{peherstorfer2016optimal} have proved that the optimization problem in Eq.\eqref{eq:opt_mf} has a unique close-form optimal solution under specific conditions on the high- and low-fidelity model. The optimal control variate coefficients are 
\begin{equation}
\beta_i^* = \rho_i \frac{\sigma_{\text{hi}}}{\sigma_i}, \quad i = 1,...,k,
\label{eq:mf_coeff}
\end{equation}
and the optimal numbers of model evaluations are 
\begin{equation}
n_0^* = \frac{C^*}{\bm{c}^T \bm{t}}, \quad n_i^* = n_0 t_i, \quad i = 1,...,k,
\label{eq:mf_number}
\end{equation}
where $\bm{t} = [1,t_1,...,t_k]^T \in \mathbb{R}^{k+1}$ are given as 
\begin{equation}
t_i = \sqrt{\frac{c_{\text{hi}} (\rho_i^2-\rho_{i+1}^2)}{c_{\text{lo}}^{(i)} (1-\rho_1^2)}}, \quad i = 1,...,k.
\label{eq:mf_t}
\end{equation}


We then compare the variance reduction performance between multifidelity estimator $\hat{s}^{mf}$ and the standard Monte Carlo estimator $\hat{s}^{mc}$ which uses the high-fidelity model along. Assuming both $\hat{s}^{mf}$ and $\hat{s}^{mc}$ have the same computational budget $C^*$, the variance reduction ratio is 
\begin{equation}
\chi = \frac{\mathbb{V}[\hat{s}^{mf}]}{\mathbb{V}[\hat{s}^{mc}]} = 
\left( \sum_{i=1}^k \sqrt{\frac{c_{\text{lo}}^{(i)}}{c_{\text{hi}}} (\rho_i^2 -\rho_{i+1}^2)} + \sqrt{1-\rho_1^2}\right)^2. \label{eq:mf_ratio}
\end{equation}
Note that the ratio $\chi$ in Eq. \eqref{eq:mf_ratio} is a sum over the correlation coefficients $\rho_1,...,\rho_k$ and the costs $c_{\text{hi}}, c_{\text{lo}}^{(1)},..., c_{\text{lo}}^{(k)}$ of all computational models in the multifidelity estimator. If variance reduction ratio $\chi <1$, the MFMC estimator is more computationally efficient than the standard Monte Carlo estimator which only uses the high-fidelity model. Eq. \eqref{eq:mf_ratio} also demonstrates that both model costs and correlation play essential roles on the efficient multifidelity estimator. The complete MFMC algorithm is summarized in {\color{black}Summary \ref{algo:MC3}} 
{\color{black}and the code implementation of MFMC can be found at \url{https://github.com/pehersto/mfmc}}.  {\color{black} Note that the step 2 and step 3 in Summary \ref{algo:MC3} are not easy to be directly estimated. Typically, the variances and correlations in such steps are estimated using samples of the various models. This leads to additional expenses, which should be contained into the total computational cost of MFMC. The estimate error of variance and correlation will have a impact on the identification of optimal control variate coefficients $\beta_i^*$ in Eq. \eqref{eq:mf_coeff} and optimal number of model evaluations ${n}_0^*$ and ${n}_i^*$ as in Eq. \eqref{eq:mf_number}.  Furthermore, the variance of the multifidelity estimator $\hat{s}^{mf}$ will increase. 
}

{\color{black}
\makeatletter
\newenvironment{megaalgorithm}[1][htb]{%
  \renewcommand{\ALG@name}{Summary}
  \begin{algorithm}[#1]%
 }{\end{algorithm}}
\makeatother

\begin{megaalgorithm}
\begin{algorithmic}[1]
\State Identify the high-fidelity model {\color{black} $\mathcal{M}_{\text{hi}}(X)$}, and low-fidelity models $\mathcal{M}_{\text{lo}}^{(1)}(X), ..., \mathcal{M}_{\text{lo}}^{(k)}(X) $
\State Determine the variances of high-fidelity model $\sigma_{\text{hi}}$ and low-fidelity models $\sigma_1,...,\sigma_k$
\State Determine the correlation coefficients $\rho_i$ of the random variable $\mathcal{M}_{\text{hi}} (X)$ stemming from the high-fidelity model and the random variables $\mathcal{M}_{\text{lo}}^{(i)} (X)$ for $ i = 1,...,k$
\State Identify the costs of the models $c_{\text{hi}}$ and $c_{\text{lo}}^{(1)},...,c_{\text{lo}}^{(k)} $ and the computational budget $C^*$
\State Set $\rho_{k+1}=0$ and define vector $\bm{t} = [1, t_1,...,t_k]^T \in \mathbb{R}_{+}^{k+1}$ as in Eq. \eqref{eq:mf_t}
\State Determine optimal control variate coefficients $\beta_i^* \in \mathbb{R}^{k+1}$ as in Eq. \eqref{eq:mf_coeff}
\State Determine optimal number of model evaluations ${n}_0^* \in \mathbb{R}_{+}$ and ${n}_i^* \in \mathbb{R}_{+}$ as in Eq. \eqref{eq:mf_number}
\State Draw $\bm{x}_1,...,\bm{x}_{n_k^*} \in \mathcal{X}$ realizations of random variable $X$
\State {\color{black}Evaluate high-fidelity model $\mathcal{M}_{\text{hi}}$ at realizations $\bm{x}_i, i=1,...,n_0^*$}
\State {\color{black}Evaluate low-fidelity models $\mathcal{M}_{\text{lo}}^{(1)}, ..., \mathcal{M}_{\text{lo}}^{(k)}$ at realizations $\bm{x}_1,...,\bm{x}_{n_l^*}$ for $l=1,...,k$}
\State Calculate the multifidelity estimator $\hat{s}^{mf}$ as in Eq. \eqref{eq:mf_1}
\end{algorithmic}
\caption{Multifidelity Monte Carlo methods for uncertainty propagation}
\label{algo:MC3}
\end{megaalgorithm}
}

\subsubsection*{\sffamily \normalsize Multifidelity Monte Carlo methods in UQ applications} 
{\color{black} MFMC has been widely applied to uncertainty quantification and propagation in engineering. }Using the standard Monte Carlo method, we often need a large number of high-fidelity model evaluations to achieve an accurate approximation of the statistical quantities. Instead, MFMC methods combining outputs from the high-fidelity models and outputs from the low-fidelity models can achieve significant cost reduction and provide unbiased estimators of the statistics. An extension of MFMC is to combine the multifidelity method with importance sampling for estimating very small probabilities of failure in reliability and risk analysis \cite{peherstorfer2016multifidelity, peherstorfer2017combining}. {\color{black} In addition, MFMC incorporated with the cross-entropy method is also employed for failure probability estimation in rare event simulation \cite{peherstorfer2018multifidelity}.} Qian et al., \cite{qian2018multifidelity} applied MFMC to present an efficient estimation of variance and sensitivity indices in the context of global sensitivity analysis, which is a particularly critical topic in the context of UQ. Gianluca et al., \cite{geraci2017multifidelity} proposed a multifidelity multilevel Monte Carlo method to accelerate uncertainty propagation (forward UQ) in aerospace applications. Fleeter et al., \cite{fleeter2019multilevel} proposed a similar hybrid method for efficient uncertainty quantification to improve the accuracy of cardiovascular hemodynamic quantities of interests given a reasonable computational cost. Jofre et al., \cite{jofre2018multi} proposed a multifidelity uncertainty quantification framework to accelerate and estimation and prediction of irradiated particle-laden turbulence simulations. {\color{black} Peherstorfer \cite{peherstorfer2019multifidelity} improved the MFMC with adaptive low-fidelity models to speed up the estimation of statistics of the high-fidelity model outputs. Quaglino et al. \cite{quaglino2019high} proposed high-dimensional and higher-order multifidelity Monte Carlo estimators, and they applied the proposed approach to a selected number of experiments, with a particular focus on cardiac electrophysiology. Fleeter et al. \cite{fleeter2019multilevel} proposed an efficient UQ framework utilizing a multilevel multifidelity Monte Carlo (MLMF) estimator to improve the accuracy of hemodynamic quantities of interest while maintaining reasonable computational cost. Gorodetsky et al. \cite{gorodetsky2020generalized} developed a generalized approximate control variate framework for multifidelity uncertainty quantification. }A recent work proposed by \cite{khan2019machine} focused on machine learning based hybrid multilevel multifidelity method, which utilizes the POD based approximation and gradient boosted tree surrogate model. 

Multifidelity methods have much broader applications, not only Monte Carlo based methods, but also more general UQ aspects, for example, optimization with uncertainty \cite{pang2017discovering, bonfiglio2018multi, heinkenschloss2018conditional}, multifidelity surrogate modeling \cite{perdikaris2015multi, parussini2017multi, giselle2019issues, guo2018analysis, chaudhuri2018multifidelity, tian2019toward} and multifidelity information reuse, and fusion \cite{cook2018generalized, perdikaris2016multifidelity}. We refer to \cite{park2017remarks, peherstorfer2018survey} for a comprehensive introduction and in-depth discussion of multifidelity methods for uncertainty propagation. 

\subsection*{\sffamily \large Multimodel Monte Carlo Methods }
In engineering practice, a common situation is to have a limited cost or time budget for data collection and thus one ends up with sparse datasets. {\color{black}This leads to epistemic uncertainty along with aleatory uncertainty, and a mix of these two sources of uncertainties (requiring imprecise probabilities \cite{augustin2014introduction}) is a particularly challenging problem. {\color{black} It has been argued that epistemic uncertainties require a different mathematical treatment than aleatory uncertainties \cite{der2009aleatory}. Arguments have been made for a variety of non-probabilistic and probabilistic treatments of epistemic uncertainties. Non-probabilistic uncertainty theories include fuzzy sets \cite{zadeh1965fuzzy}, interval methods \cite{weichselberger2000theory}, convex models \cite{ben2013convex} and Dempster-Schafer evidence theory \cite{dempster2008upper}. Probabilistic approaches include probability boxes (p-boxes) \cite{ferson2004arithmetic, dannert2020imprecise}, Bayesian \cite{sankararaman2013distribution,wei149bayesian}, random sets \cite{fetz2016imprecise, fetz2004propagation}, and frequentist theories \cite{walley1982towards}.  Walley \cite{walley1991statistical, walley2000towards} developed a unified theory of imprecise probabilities, but there are still many methods to investigate the imprecision. Beer et al. \cite{beer2013imprecise} presented an extensive review for many of these theories in engineering applications. The interested reader may find more details involving the application of imprecise probabilities in \cite{beer2013imprecise}. }}
{\color{black}
\begin{figure}[!h]
	\centering
	\includegraphics[width=6.7in]{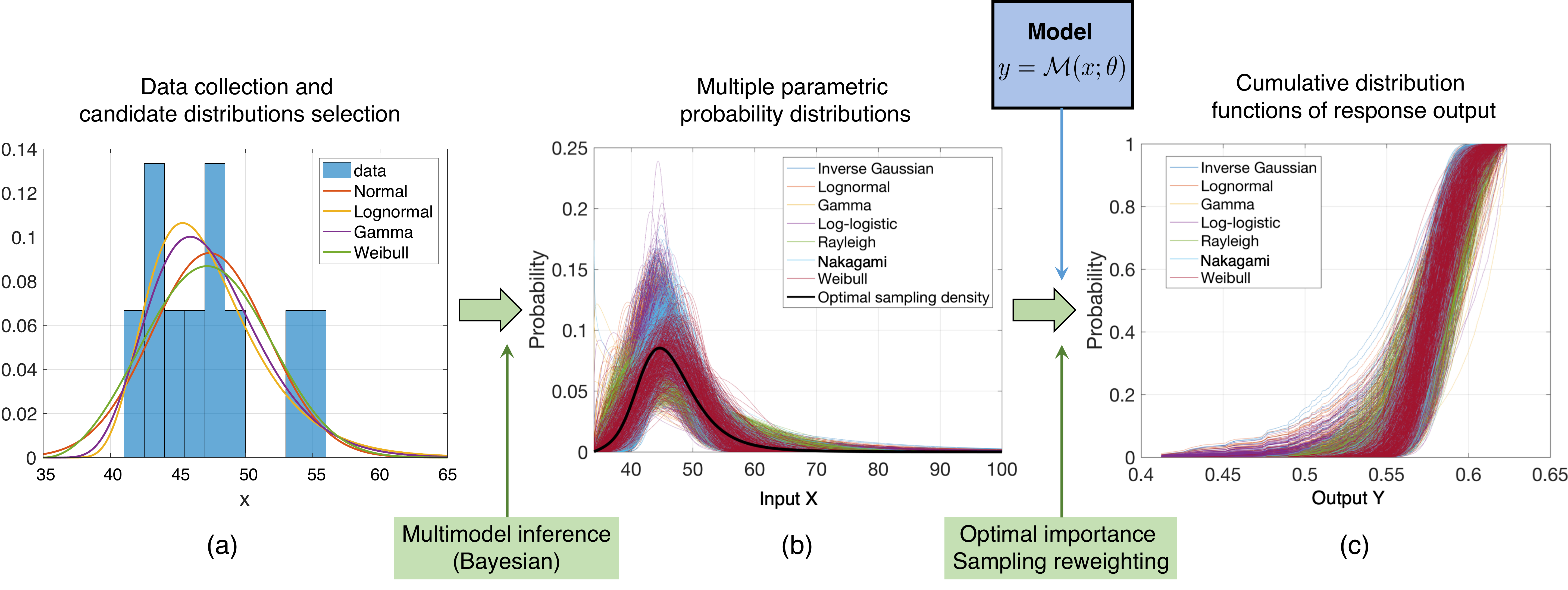}
	\caption{Illustration of multimodel Monte Carlo method. This method mainly includes (a) input data collection $X$ and candidate probability distribution selection, (b) multiple parametric probability distributions using multimodel inference and (c) propagation of uncertainties characterized by an ensemble of distributions through computational model $\mathcal{M}$ using optimal importance sampling and finally one obtains a probabilistic description of response output $Y$. }
	\label{fig2} 
\end{figure}
}

\begin{figure}[!ht]
	\centering
	\subfigure[]{\includegraphics[height=3in]{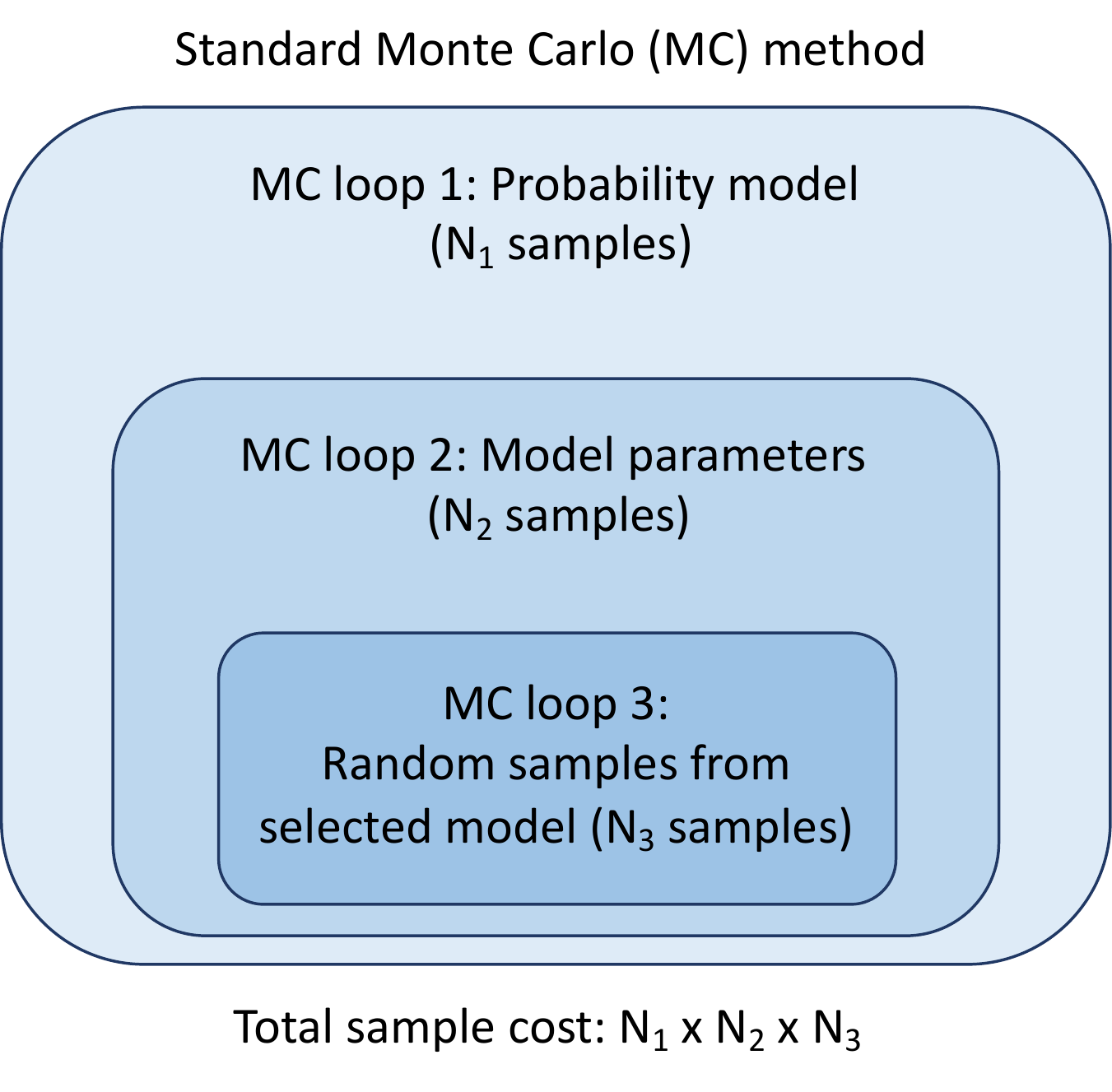}} \quad \quad \quad
	\subfigure[]{\includegraphics[height=3in]{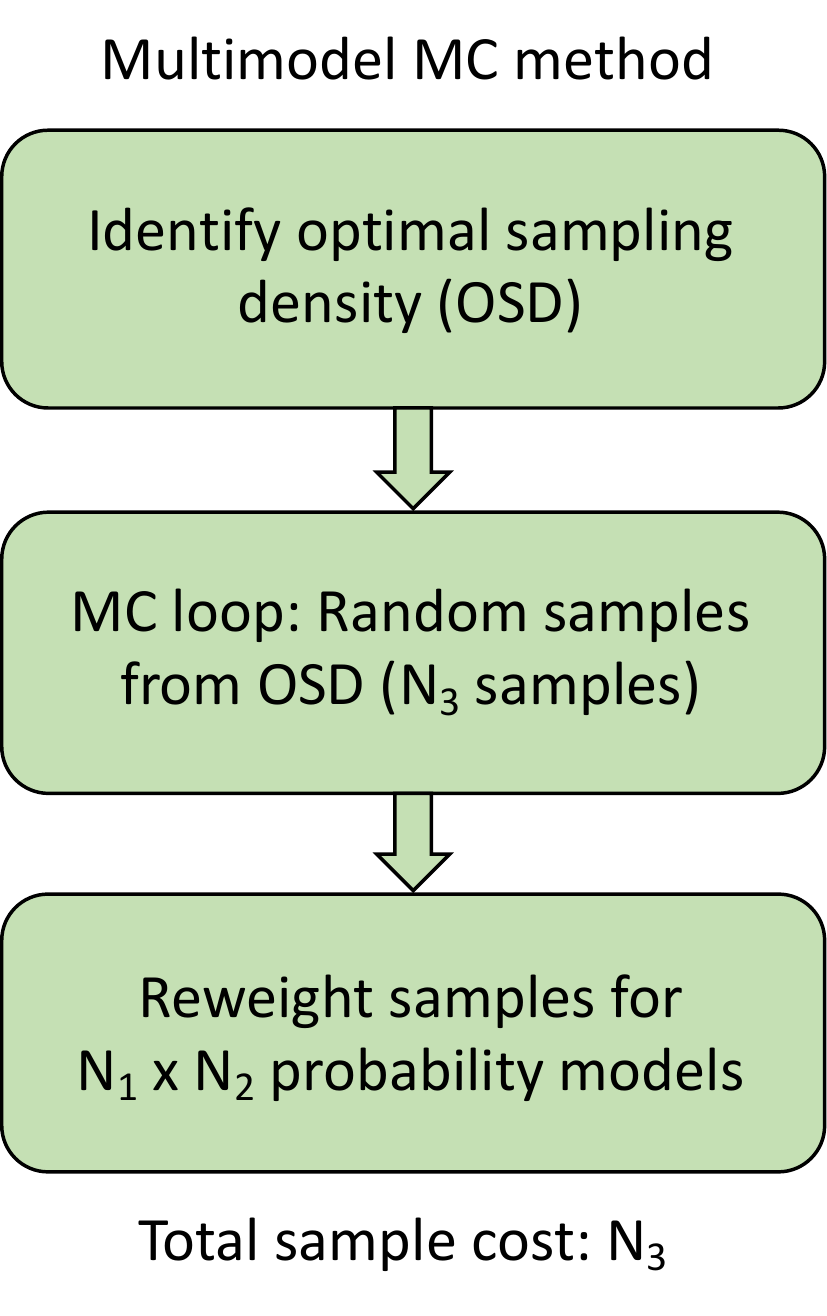}} 
	\caption{{\color{black}Conceptual comparison of (a) the standard multi-loop Monte Carlo method for propagating multiple probability models, and (b) the proposed multimodel Monte Carlo method with importance sampling reweighting}}
	\label{fig3} 
\end{figure}

Zhang and Shields \cite{zhang2018quantification} proposed a novel and efficient methodology for quantifying and propagating uncertainties resulting from a lack of data. {\color{black} As shown in Figure \ref{fig2}, the method typically starts from the input data collection (Figure \ref{fig2} (a)) and then exploits the concepts of multimodel inference from both information-theoretic and Bayesian perspectives to identify an ensemble of candidate probability distribution models (Figure \ref{fig2} (b)) and associated model probabilities that are representative of the given small datasets.} Both model-form uncertainty and model parameter uncertainty are identified and estimated within the proposed methodology. Unlike the conventional method that reduces the full probabilistic description to a single probability distribution model, the proposed method fully retains and propagates the total uncertainties quantified from all candidate probability distribution models and their model parameters through a computational model and one finally obtains a probabilistic description of response out {\color{black}(Figure \ref{fig2} (c))}. This is achieved by identifying an optimal importance sampling density that best represents the full set of probability distribution models, propagating this sampling density and reweighting the samples drawn from each of candidate probability distribution models using Monte Carlo sampling.  As a result, a complete probabilistic description of the epistemic uncertainty is achieved with several orders of magnitude reduction in Monte Carlo-based computational costs. {\color{black}As pointed out in \cite{sankararaman2013distribution}, the conventional Monte Carlo propagation of this type of uncertainties requires multiple loops, as shown in Figure \ref{fig3} (a). The proposed method provided a new direction, shown in Figure \ref{fig3} (b), to collapse these multiple loops to a single Monte Carlo loop on a surrogate distribution obtained by optimization. By propagating this surrogate distribution and reweighting the samples based on importance sampling, the proposed method achieves to simultaneously propagate uncertainty associated with a full set of probability distributions.}

Since the proposed methods integrate the multimodel inference and importance sampling for effective uncertainty quantification and propagation, we name these kinds of Monte Carlo methods as multimodel Monte Carlo method (MMMC). Note that the target of MMMC is different from the MLMC and MFMC discussed above, because existing MMMC methods focus on the quantification and propagation of input uncertainties through a {\em deterministic} computational model $\mathcal{M}$. It is possible to incorporate the stochasticity of the computational model into the MMMC framework, but this content is beyond the scope of this article.  We refer the readers to \cite{zhang2018uncertainty, zhang2018quantification} for more details and discussions.

\subsubsection*{\sffamily \normalsize Multimodel inference from small data}
Traditionally, {\color{black} model selection} is used to identify a single ``best" model given a set of candidate models and available data, the model is the sole model used for making inference from data. Any uncertainty associated with model selection is simply ignored since a single model has been selected. Nevertheless, it is difficult (and often impossible) to identify a unique best model without significant assumptions. {\color{black} For example, if very limited input data, e.g. 10 data, shown in Figure \ref{fig2} (a), is collected, it is challenging to identify a single probability distribution, e.g. Normal, Lognormal, Gamma or Weibull, to accurately represent the data.} Hence, it is necessary to consider model uncertainty and compare the validity of multiple candidate probability distribution models, which is referred to as multimodel inference, as developed by Burnham and Anderson \cite{burnham2004multimodel}.  In the multimodel inference framework, there are two approaches: information-theoretic model selection and Bayesian model selection. The information-theoretic method is implemented by establishing a criterion for the information loss resulting from approximating truth with a model. As a result, an appropriate model selection criterion is to minimize the information loss. In terms of this idea, Akaike proposed the Akaike Information Criterion (AIC) based on the fact that the expected relative {\color{black} Kullback–Leibler (K-L) divergence} could be approximated by the maximized log-likelihood function with a bias correction \cite{akaike1974new}. AIC is defined as 
{\color{black}
\begin{equation}
\text{AIC} = -2 \log(\mathscr{L}(\hat{\bm{\theta}} | \bm{d}, \psi) + 2K
\end{equation}
where $K$ is the dimension of the parameter vector $\bm{\theta}$, $\mathscr{L}(\hat{\bm{\theta}})$ is the likelihood function given the maximum likelihood estimate of the parameters $\hat{\bm{\theta}}$, $\bm{d}$ is the input data and {\color{black}$\psi$ is the probability distribution model (e.g., Normal, Lognormal, Gamma, Weibull, ect.) of input data.} It is necessary to establish a relative scale for AIC values, $\Delta_A^{(i)} = \text{AIC}^{(i)} - \text{AIC}^{\min}$ then we can estimate the distribution model probability 
\begin{equation}
\pi_i = p(\psi_i | \bm{d}) = \frac{\exp(-\frac{1}{2}\Delta_A^{(i)} )}{\sum_{i=1}^{N_p} \exp(-\frac{1}{2}\Delta_A^{(i)} ) } \label{eq:AIC}
\end{equation}
{\color{black}where $N_p$ is the number of candidate probability distribution models ${\Psi} = \left\{\psi_1, ...,\psi_{N_p}\right\}$. }

In the Bayesian setting, we consider an initial model prior probabilities $\tilde{\pi}_i = p(\psi_i)$ with $\sum_{i=1}^{N_p} \tilde{\pi}_i = 1$ for each probability distribution model $\psi_i \in {\Psi}$. Using Bayes' rule, the posterior model probability is given by 
\begin{equation}
\pi_i = p(\psi_i | \bm{d}) = \frac{p(\bm{d}|\psi_i)p(\psi_i)}{\sum_{j=1}^{N_p} p(\bm{d} |\psi_j)p(\psi_j)}, \quad i = 1,...,N_p
\label{eq:BIC}
\end{equation}  
having $\sum_{i=1}^{N_p} \pi_i = 1$ and where
\begin{equation}
p(\bm{d} | \psi_i) = \int_{\bm{\theta}_i} p(\bm{d} | \bm{\theta}_i, \psi_i) p(\bm{\theta}_i | \psi_i) d\bm{\theta}_i, \quad i = 1,..,N_p
\end{equation}
is the evidence of probability distribution model $\psi_i$. }

It is common to select the probability distribution model $\psi^* \in {\Psi}$ with the highest posterior model probability as the single ``best" model. By contrast, multimodel inference method ranks the candidate probability distribution models by their posterior model probabilities calculated by Eq. \eqref{eq:AIC} or Eq. \eqref{eq:BIC} and retains all plausible probability distribution models with non-negligible probability. {\color{black}When the plausible probability distribution models with their associated model probabilities are identified, the probability distribution model parameters (e.g., mean and standard deviation) uncertainty are assessed by Bayesian inference} 
\begin{equation}
{\color{black}
p(\bm{\theta}_i |\bm{d}, \psi_i) = \frac{p(\bm{d} | \bm{\theta}_i, \psi_i) p(\bm{\theta}_i | \psi_i)}{p(\bm{d} | \psi_i)} \propto p(\bm{d} | \bm{\theta}_i, \psi_i) p(\bm{\theta}_i | \psi_i), \quad i = 1,..,N_p. \label{eq:BI_mcmc}}
\end{equation}
The posterior $p(\bm{\theta}_i |\bm{d}, \psi_i)$ is identified implicitly through Markov chain Monte Carlo (MCMC) without requiring the calculation of model evidence $p(\bm{d} | \psi_i)$. However, the evidence is very critical in Bayesian multimodel inference and needs to be calculated with caution. We refer to \cite{zhang2018effect} for a detail discussion of the evidence calculation. 

A classical method is to identify a unique set of the probability distribution model parameters from the posterior using the maximum a posterior (MAP) estimator. However, due to a lack of data, the posterior parameter probability will likely possess a large variance. Thus, we retain the full posterior densities for each plausible probability distribution model instead of discarding the full uncertainty by selecting a single set of MAP estimator or integrating out its variability using model averaging methods. Theoretically, combining model-form uncertainty and model parameter uncertainty yields an infinite set of parametrized probability distribution models. Practically, it is necessary to reduce this to a finite but statistically representative set of models. {\color{black}This is achieved by Monte Carlo sampling, which randomly selects a probability distribution model family from ${\Psi}$ with model probabilities $\pi_i$ and randomly selects its parameters from the joint parameter densities $p(\bm{\theta}_i |\bm{d}, \psi_i) $}. A detailed discussion of this process can be found in \cite{zhang2018quantification}. {\color{black}Now we have a large number of plausible probability distribution models, and the key question is how to efficiently propagate these input probability distributions through a computational model $\mathcal{M}$. }

\subsubsection*{\sffamily \normalsize Importance sampling}
Importance sampling involves a change of probability measure. Instead of taking ${X}$ from a distribution with $p$, we draw random samples $\bm{x}_1,...,\bm{x}_n \in \mathcal{X}$ from an alternative pdf $q$ to estimate the expectation
\begin{equation}
\mu_q = \mathbb{E}_q[\mathcal{M}] = \int_{\mathcal{X}}\mathcal{M}(\bm{x}) \frac{p(\bm{x})}{q(\bm{x})} q(\bm{x}) d\bm{x} = \frac{1}{n} \sum_{i=1}^n \mathcal{M}(\bm{x}_i)\frac{p(\bm{x}_i)}{q(\bm{x}_i)} = \frac{1}{n} \sum_{i=1}^n \mathcal{M}(\bm{x}_i)w(\bm{x}_i)
\end{equation}
where $\mathbb{E}_q[\cdot]$ is the expectation with respect to $q(\bm{x})$ and $\mathcal{M}$ is the computational model. The ratios $w(\bm{x}) = p(\bm{x})/q(\bm{x})$, as the importance weights, play a fundamental role in the importance sampling estimator. 

The importance sampling method is also an unbiased estimator, which means that $\mathbb{E}[\mu_q] = \mathbb{E}[\mathcal{M}]$. We are interested to study the variance of $\mu_q$, which is $\mathbb{V}_q[\mu_q] = \sigma_q^2/n$, where 
\begin{equation}
\sigma_q^2 = \int_{\mathcal{X}} \frac{(\mathcal{M}(\bm{x})p(\bm{x}))^2}{q(\bm{x})}d\bm{x} - \mu^2 = \int_{\mathcal{X}} \frac{(\mathcal{M}(\bm{x})p(\bm{x}) - \mu q(\bm{x}))^2}{q(\bm{x})}d\bm{x}. 
\label{eq:is_var}
\end{equation} 
where $\mu =\mathbb{E}[\mathcal{M}]$ is the true estimator. 

From the second expression in Eq. \eqref{eq:is_var}, we note that 
\begin{equation}
\sigma_q^2 = \mathbb{E}_q[ (\mathcal{M}({X})p({X})-\mu q({X}))^2 / q({X})^2].
\end{equation}

The variance estimate is therefore written as 
\begin{equation}
\tilde { \sigma}_q^2 = \frac{1}{n}\sum_{i=1}^n \left( \frac{\mathcal{M}(\bm{x}_i)p(\bm{x}_i)}{q(\bm{x}_i)} -\mu_q \right)^2 = \frac{1}{n}\sum_{i=1}^n \left(w(\bm{x}_i) \mathcal{M}(\bm{x}_i) - \mu_q \right)^2. 
\end{equation}

The variance estimate above guides us how to select a good sampling density $q(\bm{x})$ to reduce variance in the importance sampling. Assume that $\mathcal{M}(\bm{x}) \ge 0$ and $\mu>0$, then the optimal sampling density is given by 
\begin{equation}
q^*(\bm{x}) = \frac{\mathcal{M}(\bm{x})p(\bm{x})}{\mathbb{E}[\mathcal{M}(\bm{x})]} = \frac{\mathcal{M}(\bm{x})p(\bm{x})}{\mu} 
\end{equation}
which achieves $\sigma_q^2=0$ but is always infeasible in practice. This is because that we could compute $\mu$ directly from $\mathcal{M}, p$ and $q$ without any sampling \cite{mcbook}.

\subsubsection*{\sffamily \normalsize Uncertainty propagation using optimal importance sampling}
Instead of achieving a variance reduction, we are more interested in ensuring that our sampling density is as \textcolor{black}{close} as possible to the target density $p(\bm{x})$, given the difficulty of sampling from $p(\bm{x})$ itself. This is achieved by minimizing the $f$-divergence which defines the difference between two distributions $P$ and $Q$ over a space $\Omega$ with measure $\mu$ as:
\begin{equation}
D_f(P\parallel Q)=\int_{\Omega}f\left(\frac{p(\bm{x})}{q(\bm{x})}\right)q(\bm{x})d\mu(\bm{x}) \label{eq:divergence}
\end{equation} 
Various functions $f(\cdot)$ have been proposed based on the basic definition in Eq.\ (\ref{eq:divergence}), for example, Kullback-Leibler divergence, Hellinger distance, total variation distance and mean square difference. 


Zhang and Shields \cite{zhang2018quantification} provided an explicit analytical derivation for the optimal importance sampling density given an ensemble of candidate target probability densities. The approach firstly introduced a widely used metric, the mean square difference (MSD), to quantify the difference between one importance sampling density and one target probability density, which is given by:
\begin{equation}
{\mathcal{H}}(P\parallel Q)={ \frac { 1 }{ 2 } \int{ \left( { p(\bm{x} |\bm{\theta}) } - { q(\bm{x}) } \right) ^{ 2 }d\bm{x} } } \label{eq: MSD} .
\end{equation}
{\color{black}
The corresponding total expected mean squared difference between a single sampling density $q(\bm{x})$ and the ensemble of ${N_p}$ probability target densities $p_j(\bm{x}|\bm{\theta}_j, \psi_j), j=1,...,{N_p}$ can be formulated as:
\begin{equation}
\mathscr{E} = \sum_{j=1}^{{N_p}}E\left[{\mathcal{H}}(\mathcal{P}_j \parallel Q)\right] = {E_{\theta} }\left[ \int{\sum _{ j=1 }^{ {{N_{\color{black}p}}}} \frac{1}{2} { \left( { p_j(\bm{x} | \bm{\theta}_j, \psi_j) } - { q(\bm{x}) } \right) ^{ 2 }} d\bm{x}} \right]
\label{eq: total_EMSD} 
\end{equation}
}
To ensure the sampling density $q(\bm{x})$ is as close as possible to the multiple target probability densities $p_j(\bm{x}|\bm{\theta}_j, \psi_j), j=1,...,{N_p}$, an overall optimization problem is solved to minimize the total expected mean squared difference expressed as a functional ${\mathcal{L}}(q)$ given isoperimetric constraint ${\mathcal{I}}(q)$
\begin{equation}
\begin{aligned}
& \underset{q}{\text{minimize}}
& &{\mathcal{L}}(q)=E_{\theta}\left [ \int{{{\mathcal{F}} }(\bm{x}, \bm{\theta}, q(\bm{x}))}d\bm{x} \right] \\
& \text{subject to}
& &{\mathcal{I}}(q) = \int{q(\bm{x})d\bm{x}}-1=0 \label{eq: opt_EMSD}
\end{aligned}
\end{equation}
where the action functional $\mathcal{F}(\cdot)$ is \textcolor{black}{the total square differences}:
\begin{equation}
{\color{black}
{\mathcal{F}(\bm{x}, \bm{\theta}, q(\bm{x}))}={ \frac { 1 }{ 2 } \sum_{j=1}^{{N_{p}}}{ \left( { p_j(\bm{x} | \bm{\theta}_j,\psi_j) } - { q(\bm{x}) } \right) ^{ 2 }} } \label{eq:MSD_funcitonal} 
}
\end{equation}
\textcolor{black}{ and $E_{\theta}$ is the expectation with respect to the posterior probability of the model parameters $\bm{\theta}$.} ${\mathcal{I}}(q)$ ensures that $q(\bm{x})$ is a valid probability density function. Notice that the optimization problem in Eq.\eqref{eq: opt_EMSD} has a closed-form solution given by the convex mixture model \cite{zhang2018quantification}
\begin{equation}
{\color{black}
q^*(\bm{x}) =\frac{1}{{N_{\color{black}p}}} \sum_{j=1}^{{N_{\color{black}p}}}E_{\theta}\left[{p_j({\bm{x} | \bm{\theta}_j, \psi_j})}\right] \label{eq: opt_MSD2}
}
\end{equation}
and this solution can be generalized to combine the posterior model probabilities as 
\begin{equation}
{\color{black}
q^*(\bm{x}) = \sum_{j=1}^{{N_{\color{black}p}}} {\color{black}{\pi}_j} E_{\theta} \left [ { p_j(\bm{x}|\bm{\theta}_j, \psi_j)} \right] \label{eq: opt_MSD3}
}
\end{equation}
where $ \pi_j$ is the posterior model probability for model $\psi_j$, computed by Eq. \eqref{eq:AIC} or Eq. \eqref{eq:BIC}. 

Samples are drawn from $q^*(\bm{x})$ are re-weighted based on the importance weights. In other words, each sample drawn from $q^*(\bm{x})$ is re-weighted a large number of times according to each plausible probability distribution model. We thus simultaneously propagate an ensemble of probability distribution models and achieve a significant improvement, which is to reduce a multi-loop Monte Carlo with $n^3$ samples to a single loop Monte Carlo with $n$ samples, as shown in Figure \ref{fig3}. Moreover, the developed method provides a high degree of flexibility, and consequently, it is easy and adaptively updated to accommodate additional new collected data or new candidate probability distribution models but without additional computational cost. We added a summary of the MMMC algorithm procedure as {\color{black}Summary \ref{algo:MC4}} and we also refer the interested readers to \cite{zhang2018quantification, zhang2019efficient} for in-depth analysis and discussion. {\color{black} The code implementation of MMMC (e.g., multimodel inference and importance sampling) can be found at {\textit{UQpy}} (\url{https://github.com/SURGroup/UQpy}). {\textit{UQpy}} (Uncertainty Quantification with Python) is an open-source Python package for general UQ in mathematical and physical systems and it serves as both a user-ready toolbox that includes many of the latest methods for UQ in computational modeling and a convenient development environment for Python programmers advancing the field of UQ \cite{uqpy}}.



\makeatletter
\newenvironment{megaalgorithm}[1][htb]{%
  \renewcommand{\ALG@name}{Summary}
  \begin{algorithm}[#1]%
 }{\end{algorithm}}
\makeatother

\begin{megaalgorithm}
\begin{algorithmic}[1]
\State Collected initial limited data $\bm{d}$ from experiments or simulations
\State {\color{black}Identify candidate probability distribution models {\color{black}$\Psi = \left\{ \psi_1, ...,\psi_{N_p}\right\}$ for input data $\bm{d}$}}
\State Compute model probabilities using information-theoretic in Eq. \eqref{eq:AIC} or Bayesian multimodel inference in Eq. \eqref{eq:BIC}
\State {\color{black}Estimate the posterior joint parameter density $p(\bm{\theta} | \bm{d}, \psi_i)$ for each plausible probability distribution model $\psi_i$ using Bayesian inference with MCMC, as in Eq. \eqref{eq:BI_mcmc}}
\State Establish a finite model set by randomly selecting the model family $\psi_i$ with model probability $\pi_i$ and randomly generating parameter values from $p(\bm{\theta} | \bm{d}, \psi_i)$ 
\State Determine the optimal sampling density $q^{*}(\bm{x})$ by solving the optimization problem defined in Eq \eqref{eq: opt_EMSD}
\State {\color{black}Draw random samples from $q^{*}(\bm{x})$ and evaluate the computational model $\mathcal{M}(\bm{x})$}
\State Propagate the uncertainties by reweighting the samples according to the importance weights $w(\bm{x}) = p(\bm{x})/q(\bm{x})$
\State Compute the statistical estimator, e.g. mean of the response output
\end{algorithmic}
\caption{Multimodel Monte Carlo methods for uncertainty propagation}
\label{algo:MC4}
\end{megaalgorithm}

\subsubsection*{\sffamily \normalsize Multimodel Monte Carlo extensions and applications}  
{\color{black} Along with the MMMC framework, Zhang and Shields further investigated the effect of prior probability on quantification and propagation of imprecise probabilities resulting from small datasets \cite{zhang2018effect}. It is demonstrated that prior probabilities play a critical role in Bayesian multimodel UQ framework for small datasets, and inappropriate priors may lead to biased probabilities as well as inaccurate estimators even for large datasets. When a multidimensional UQ problem is involved, a further study generalizes this MMMC methodology to overcome the limitations of the independence assumption (or subjective Gaussian correlation assumption) by introducing a flexible copula dependence model to capture complex dependencies \cite{zhangcopula}. Zhang et al. also extended the MMMC framework to integrate uncertainties into sensitivity index estimators and proposed an imprecise global sensitivity analysis method \cite{zhangigsa}. This method provides a full probabilistic description of Sobol' indices, whose distribution characterizes uncertainty in the sensitivity resulting from small dataset size. The proposed method has been applied to many real-world science and engineering problems, for example, material science \cite{bostanabad2018uncertainty, zhangmams, zhang2020robust,zhangcms2020}, structural reliability \cite{sundar2019reliability, wang2018computing, liu2018novel, sofi2020propagation,songstochastic}, failure and risk assessment \cite{guo2019imprecise, manouchehrynia2020fatigue, wang2020augmented}, etc. 

Based on the MMMC framework, Gao et al. \cite{gao2019nonparametric} proposed a nonparametric-based approach for the characterization and propagation of epistemic uncertainty due to small datasets. Peng et al. \cite{peng2018nonparametric} developed a nonparametric uncertainty representation method with different insufficient data from two sources. Troffaes \cite{troffaes2018imprecise} further proposed an imprecise Monte Carlo simulation and iterative importance sampling for the estimation of lower previsions. Fetz \cite{fetz2019improving} improved the convergence of iterative importance sampling for computing upper and lower expectations. Decadt et al. \cite{decadt2019monte} proposed to investigate Monte Carlo methods for estimating lower envelopes of expectations of real random variables. Wei et al. \cite{wei2019non} developed a non-intrusive stochastic analysis with parameterized imprecise probability models. Aakash et al. \cite{satish2017probabilistic} applied MMMC to investigate probabilistic calibration of material models from limited data and its influence on structural response during fires. Song et al. \cite{song2019generalization,songstochastic} generalized to propose a non-intrusive imprecise stochastic simulation for mixed uncertain variables in the NASA Langley UQ challenge problem. 
}

\section*{\sffamily \Large \MakeUppercase{Conclusions}}
 In many cases across computational science and engineering, uncertainty quantification is playing an increasingly important role in computationally evaluating the performance of complex mathematical, physical, and engineering systems. Typically, a computationally expensive high-fidelity model characterizes the system with high accuracy but high costs. Thus the standard Monte Carlo method is often very time consuming because it relies on a large number of random samples (model evaluations) to estimate the statistical quantities of response outputs. Several efficient Monte Carlo methods are therefore proposed to address the computational challenges. Multilevel Monte Carlo method (MLMC) utilizes control variates technique to reduce the computational cost by performing most simulations at a relatively low cost and only a few simulations at a high-cost. Similar to the MLMC method, the multifidelity Monte Carlo method (MFMC), as a variant of the control variates, aims to combine high-fidelity models and low-fidelity models to speed up the statistical estimation. In the context of imprecise probabilities, typically arising from small data issues, the multimodel Monte Carlo method (MMMC) is developed to quantify the uncertainties using multimodel inference, which combines the model-form and model parameter uncertainties, and then efficiently propagate an ensemble of probability models through the optimal importance sampling reweighting scheme. {\color{black} These efficient modern Monte Carlo methods can be employed to address many UQ challenges, not only for forward UQ problems, but also more general UQ related issues, e.g., optimization with uncertainty, robust design with uncertainty, and UQ in artificial intelligence and machine learning. }

\section*{\sffamily \Large ACKNOWLEDGEMENTS}
The author is grateful to the reviewers for their insightful and helpful comments and suggestions on the earlier version of the manuscript. This material was based upon work supported by the U.S. Department of Energy, Office of Science, Office of Advanced Scientific Computing Research, Applied Mathematics program under contract and award numbers ERKJ352, ERKJ369, and by the Artificial Intelligence Initiative at the Oak Ridge National Laboratory (ORNL). ORNL is operated by UT Battelle, LLC., for the U.S. Department of Energy under Contract DE-AC05-00OR22725.

\section*{\sffamily \Large \MakeUppercase{Conflict of Interest}}
The author has declared no conflicts of interest for this article. 
%
%
%
%
%
%
\bibliography{references.bib}
%

\end{document}